\documentstyle[eqsecnum,prd,aps,epsf]{revtex}
\begin{document}
\draft
\title{Correlation Functions of CMB Anisotropy and Polarization}
\author{Kin-Wang Ng\footnote{\tt nkw@phys.sinica.edu.tw} 
and Guo-Chin Liu\footnote{\tt liugc@phys.sinica.edu.tw}} 
\address{Institute of Physics, Academia Sinica, Taipei, Taiwan 11529, R.O.C.}
\maketitle

\begin{abstract}
We give a full analysis of the auto- and cross-correlations between the 
Stokes parameters of the cosmic microwave background. In particular, 
we derive the windowing function for an antenna with Gaussian response in
polarization experiment, and construct correlation function estimators
corrected for instrumental noise. They are applied to calculate the
signal to noise ratios for future anisotropy and polarization
measurements. While the small-angular-scale anisotropy-polarization 
correlation would be likely detected by the MAP satellite, 
the detection of electric and magnetic polarization would require higher
experimental sensitivity. For large-angular-scale measurements such
as the being planned SPOrt/ISS, the expected signal to noise ratio
for polarization is greater than one only for reionized models with high 
reionization redshifts, and the ratio is less 
for anisotropy-polarization correlation. Correlation and covariance matrices
for likelihood analyses of ground-based and satellite data are also given.
\end{abstract}


\section{Introduction}

The detection of the large-angle anisotropy of the cosmic microwave
background (CMB) by the {\it COBE} DMR experiment~\cite{smo} provided
important evidence of large-scale spacetime inhomogenities. Since then, 
a dozen of small-scale anisotropy measurements have hinted that 
the Doppler peak resulting from acoustic oscillations of the baryon-photon
plasma on the last scattering surface seems to be present~\cite{pag}.
CMB measurements gain an advantage over other traditional observations
due to the fact that the small CMB fluctuations can be well treated as linear,
while the low-redshift universe is in a non-linear regime. 
It is now well established that CMB temperature anisotropies are genuine 
imprint of the early universe, which could potentially be used to determine
to a high precision virtually all cosmological parameters of interest.
It has been estimated that a number of cosmological parameters can
be determined with standard errors of $10\%$ or better by the upcoming NASA
MAP satellite~\cite{jun}. Furthermore, the future Planck Surveyor CMB mission
would have capability of observing the early universe about 100 times better
than MAP.

At this point, we should explore as much information as possible besides
the temperature anisotropy contained in the relic photons. 
Anisotropic radiation possessing a non-zero quadrupole moment 
acquires a net linear polarization when it is scattered with electrons via
Thomson scattering~\cite{ree} (also see Eq.~(6) of Ref.~\cite{ng3}). 
When the photons begin to decouple from the matter on the last scattering
surface and develope a quadrupole anisotropy 
via Sachs-Wolfe effect~\cite{sac},
linear polarization is created from scatterings with free electrons near
the last scattering surface. Studies have shown that on small angular scales 
the rms polarization, in a standard universe, is a few percents of the rms
anisotropy, while the large-scale polarization is insignificant~\cite{bon}. 
In models with early reionization, the large-scale
polarization is greatly enhanced, to a few percents level, 
but the small-scale anisotropy is suppressed significantly~\cite{ng3,zal1}. 
Therefore, CMB polarization would provide
a valuable complementary information to the anisotropy measurements. 
In addition, the anisotropy-polarization cross correlation offers a test of
physics on the last scattering surface, as well as a possibility of
distinguishing the scalar and tensor perturbations~\cite{cri1,cri2}. 
However, all of these polarization calculations have relied on 
a small-angle approximation, which may not be valid when a large 
sky-coverage is considered.

As such, full-sky analyses of the polarization have been performed~
\cite{ng4,sel1,kam1,sel2,kam2}. 
It was found that there are modifications to 
low multipole moments ($l<30$) of the polarization power spectra, where the
tensor contribution dominates over the scalar contribution~\cite{ng4,sel2}. 
More importantly, rotationally invariant power spectra of the Stokes 
parameters have been constructed~\cite{sel1,kam1,sel2,kam2}.
In particular, one of them is a parity-odd magnetic polarization spectrum,
which vanishes for scalar-induced polarization, thereby allowing one to make 
a model-independent identification of non-scalar (i.e. vector or tensor) 
perturbations (also see Ref.~\cite{sel}). 
Recently it was shown that magnetic polarization would be a strong
discriminator between defect and inflation models~\cite{hu1,sel3}.
Also, a new physically transparent formalism based on the total angular
momentum representation~\cite{tol} was proposed~\cite{hu1,hu2}, 
which simplifies the radiative transport problem and can be easily 
generalized to open universes~\cite{hu3}.

Since polarization fluctuations are typically at a part in a million, an
order magnitude below the temperature fluctuations, to measure this signal
requires high detector sensitivity, long integration time, and/or a large
number of pixels. So far, only experimental upper limits have been 
obtained~\cite{lub,par,net}, 
with the current limit on the linear polarization being $16\mu K$~\cite{net}.
Ground-based experiments being planned or built
will probably achieve detection sensitivity using low-noise HEMT amplifiers
as well as long hours of integration time per pixel. The MAP satellite will 
launch in 2000 and make polarization measurements of the whole sky in about
$10^5$ pixels. If the polarization foreground can be successfully removed,
MAP should marginally reach the detection level. For a detection of the 
magnetic polarization one would require either several years of MAP 
observations or the Planck mission~\cite{sel1,sel3,kam3}. We expect that
polarization measurements are as important as anisotropy in future missions. 
 
Previous full-sky studies of the polarization 
are mainly based on angular power
spectrum estimators in Fourier space. Although electric- and magnetic-type 
scalar fields $E$ and $B$ in real space can be constructed,
they must involve nonlocal derivatives of the Stokes components.
In this paper, we will study in detail the auto- and cross-correlation 
functions of the Stokes parameters themselves in real space. 
Although the two approaches should be equivalent to each other, one can find
individual advantages in different situations.     
We will follow the formalism of Ref.~\cite{sel1}, expanding the Stokes
parameters in terms of spin-weighted spherical harmonics. The expansion
coefficients are rotationally invariant power spectra which will be evaluated
using CMBFAST Boltzmann code developed by Seljak and Zaldarriaga~\cite{zal2}. 
In Sec.~\ref{stokes} we briefly introduce the
CMB Stokes parameters and their relation to spin-weighted spherical
harmonics. Sec.~\ref{spin} is devoted to discussions of the properties of
the harmonics, the harmonics representation of rotation group, and
the generalized addition theorem and recursion relation.
In Sec.~\ref{power} we expand the Stokes parameters in spin-weighted 
harmonics, and briefly explain how to compute the power spectra
induced by scalar and tensor perturbations. In Sec.~\ref{window} we derive
window functions appropriate to detectors
with Gaussian angular response in anisotropy and polarization experiments.
The instrumental noise of detectors in CMB measurements is treated in 
Sec.~\ref{noise} as white noise superposed upon the microwave sky.
Sec.~\ref{estimator} is to construct the auto- and
cross-correlation function estimators corrected for noise bias
in terms of the power spectra. 
As examples, in Sec.~\ref{result} we compute the means 
and variances of the estimators for different configurations
of future space missions in standard cold dark matter models. 
Further, we outline the 
likelihood analysis of the experimental data in Sec.~\ref{like}.
Sec.~\ref{conclusion} is our conclusions.

\section{Stokes Parameters}\label{stokes}

\begin{figure}
\leavevmode
\hbox{
\epsfxsize=6.5in
\epsffile{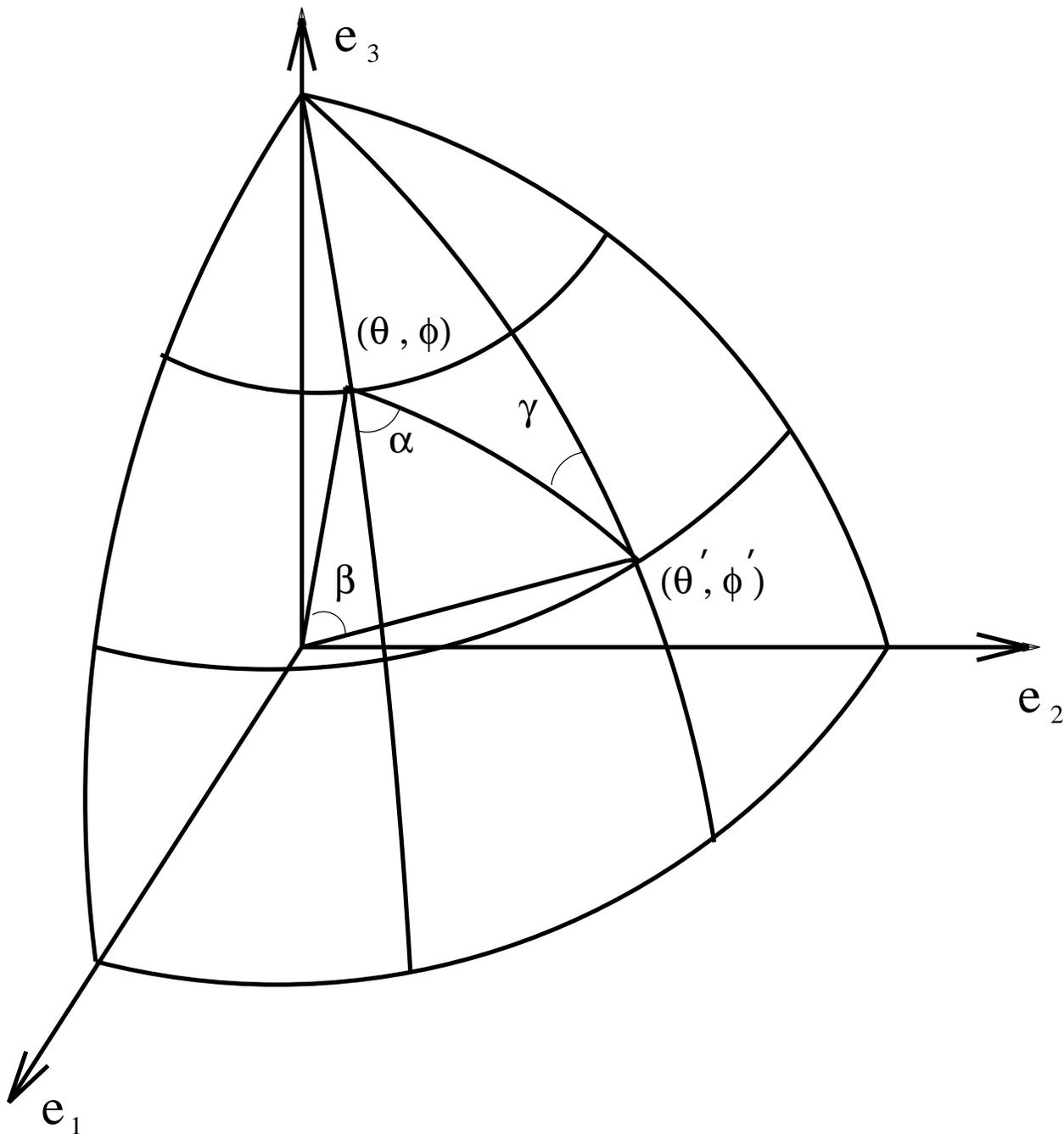}}
\caption{Spherical coordinates showing two unit vectors 
$\hat n'(\theta',\phi')$ and $\hat n(\theta,\phi)$ with separation angle 
$\beta$. The angles between the great arc connecting the two points and
the longitudes are $\gamma$ and $\alpha$.}
\label{fig1}
\end{figure}

Polarized light is conventionally described in terms of the four Stokes
parameters $(I,Q,U,V)$, where $I$ is the intensity, $Q$ and $U$ represent 
the linear polarization, and $V$ describes the circular polarization.
Each parameter is a function of the photon propogation direction $\hat n$. 
Let us define 
\begin{equation}
T=I-{\bar I}
\end{equation}
as the temperature fluctuation about the mean. Since circular polarization 
cannot be generated by Thomson scattering alone, $V$ decouples from the other
components. So, it suffices to consider only the Stokes components $(T,Q,U)$
as far as CMB anisotropy and polarization is concerned. Traditionally, for
radiation propagating radially along $\hat e_r$ in the spherical coordinate
system, see Fig.~\ref{fig1}, $Q$ and $U$ are defined with respect to an
orthonormal basis $(\hat a, \hat b)$ on the sphere, 
which are related to $(\hat e_\theta, \hat e_\phi)$ by
\begin{equation}
\hat a = \hat e_\phi,\quad{\rm and}\quad \hat b = - \hat e_\theta.
\end{equation}
Then, $Q$ is the difference in intensity polarized in the $\hat b$ and
$\hat a$ directions, while $U$ is the difference in
the $(\hat a + \hat b)/\sqrt 2$ and $(\hat a - \hat b)/\sqrt 2$
directions~\cite{cha}. 
Under a left-handed rotation of the basis about $\hat e_r$  
through an angle $\psi$,
\begin{equation}
\left ( \begin{array}{c} \hat a'\\ \hat b' \end{array} \right)=
\left( \begin{array}{cc} 
\cos\psi&-\sin\psi\\ \sin\psi&\cos\psi
\end{array} \right)
\left ( \begin{array}{c} \hat a\\ \hat b \end{array} \right),
\end{equation}
or equivalently,
\begin{equation}
\frac{1}{\sqrt 2} \left(\hat a'+i\hat b'\right) = e^{i\psi}
\frac{1}{\sqrt 2} \left(\hat a+i\hat b\right).
\label{rotation}
\end{equation} 
Under this transformation $T$ and $V$ are invariant while 
$Q$ and $U$ being transformed to~\cite{cha}
\begin{equation}
\left ( \begin{array}{c} Q'\\U' \end{array} \right)=
\left( \begin{array}{cc} 
\cos2\psi&\sin2\psi\\-\sin2\psi&\cos2\psi
\end{array} \right)
\left ( \begin{array}{c} Q\\U \end{array} \right),
\end{equation}
which in complex form is
\begin{equation}
Q'(\hat e_r)\pm i U'(\hat e_r) = e^{\mp 2i\psi}
\left[ Q(\hat e_r)\pm i U(\hat e_r) \right].
\end{equation}
Hence, $Q(\hat n)\pm i U(\hat n)$ has spin-weight $\mp 2$.\footnote 
{Generally, a quantity $\eta$ will be said to have spin-weight $s$ if it
transforms as $\eta'=e^{s i\psi}\eta$ under the
rotation~(\ref{rotation})~\cite{new}.}
Therefore, we may expand each Stokes parameter in its appropriate 
spin-weighted spherical harmonics~\cite{tol,sel1}. 

Unfortunately, the convention in theory has a little difference from 
experimental practice. 
In CMB polarization measurements, usually the north celestial pole is 
chosen as the reference axis $\hat e_3$, and linear polarization 
at a point $\hat x$ on the celestial sphere is defined by
\begin{equation}
{\cal Q}(\hat x)=T_{N,S}-T_{E,W},\quad{\rm and}\quad 
{\cal U}(\hat x)=T_{NE,SW}-T_{NW,SE},
\end{equation}
where $T_{N,S}$ is the antenna temperature of radiation polarized along 
the north-south direction, and so on~\cite{lub}.
In small-scale experiments covering only small patches of the sky, 
the geometry is essentially flat, so one can simply choose any local 
rectangular coordinates to define ${\cal Q}$ and ${\cal U}$.
Since an observation in direction $\hat x$ receives radiation with
propagating direction $\hat n = -\hat x$, we have
\begin{equation}
{\cal Q}(\hat x) = Q(\hat n),\quad{\rm and}\quad {\cal U}(\hat x) 
= -U(\hat n).
\end{equation}

\section{Spin-weighted Spherical Harmonics}\label{spin}
An explicit expression of spin-$s$ spherical harmonics is~\footnote
{In Ref.~\cite{new}, the sign $(-1)^m$ is absent. 
We have added the sign in order to match the conventional 
definition for $Y_{lm}$.}
\cite{new,pen}
\begin{eqnarray}
\:_{s}Y_{lm}(\theta,\phi)&&=(-1)^{m}e^{im\phi}\left[\frac{2l+1}{4\pi}
\frac{(l+m)!}{(l+s)!}\frac{(l-m)!}{(l-s)!}\right]^{\frac{1}{2}}
\sin^{2l} \left(\frac{\theta}{2}\right) \nonumber \\
&&\times\sum_{r}\left(\begin{array}{c}
l-s\\
r\\
\end{array}\right)\left(\begin{array}{c}
l+s\\
r+s-m\\
\end{array}\right)(-1)^{l-s-r}\cot^{2r+s-m}\left(\frac{\theta}{2}\right),
\label{s-har}
\end{eqnarray}
where
\begin{equation}
\max(0,m-s) \le r\le \min(l-s,l+m).
\end{equation}
Note that the common spherical harmonics $Y_{lm}=\:_{0}Y_{lm}$. 
They have the conjugation relation and parity relation:
\begin{equation}
\:_{s}Y^*_{lm}(\theta,\phi)=(-1)^{m+s}\:_{-s}Y_{l-m}(\theta,\phi),
\label{conjugation}
\end{equation}
\begin{equation}
\:_{s}Y_{lm}(\pi-\theta,\phi+\pi)=(-1)^{l}\:_{-s}Y_{lm}(\theta,\phi).
\label{parity}
\end{equation}
They satisfy the orthonormality condition and completeness relation:
\begin{equation}
\int d\Omega \:_{s}Y^*_{l'm'}(\theta,\phi)\:_{s}Y_{lm}(\theta,\phi)
=\delta_{l'l}\delta_{m'm},
\label{ortho}
\end{equation}
\begin{equation}
\sum_{lm}\:_{s}Y^*_{lm}(\theta',\phi')\:_{s}Y_{lm}(\theta,\phi)
=\delta(\phi'-\phi) \delta(\cos\theta'-\cos\theta).
\end{equation}
Therefore, a quantity $\eta$ of spin-weight $s$ defined on the sphere 
can be expanded in spin-$s$ basis,
\begin{equation}
\eta(\theta,\phi)=\sum_{lm} \eta_{lm} \:_{s}Y_{lm} (\theta,\phi),
\end{equation}
where the expansion coefficients $\eta_{lm}$ are scalars. 

The raising and lowering operators, $\partial\!\!'$ and 
$\bar{\partial\!\!'}$,
acting on $\eta$ of spin-weight $s$, are defined by~\cite{new}
\begin{eqnarray}
{\partial\!\!'}\eta&=&-(\sin\theta)^s \left[\frac{\partial}{\partial\theta}
  +i\csc\theta\frac{\partial}{\partial\phi}\right](\sin\theta)^{-s}\eta,\\
\bar{\partial\!\!'}\eta&=&-(\sin\theta)^{-s}\left[\frac{\partial}{\partial
  \theta} -i\csc\theta\frac{\partial}{\partial\phi}\right](\sin\theta)^s\eta.
\end{eqnarray}
When they act on the spin-$s$ spherical harmonics, we have~\cite{new}
\begin{eqnarray}
{\partial\!\!'}\:_{s}Y_{lm}&=&\left[(l-s)(l+s+1)\right]^{1\over 2} 
                            \:_{s+1}Y_{lm},\\
\bar{\partial\!\!'}\:_{s}Y_{lm}&=&-\left[(l+s)(l-s+1)\right]^{1\over 2} 
                            \:_{s-1}Y_{lm},\\
\bar{\partial\!\!'}{\partial\!\!'}\:_{s}Y_{lm}&=&-(l-s)(l+s+1)\:_{s}Y_{lm}.
\end{eqnarray}
Using these raising and lowering operations, we obtain the generalized 
recursion relation for $l-2\ge \max(|s|,|m|)$, 
\begin{eqnarray}
\left(\frac{l+s}{l-s}\right)^{1\over 2}
&&\:_{s}Y_{lm} =\left[\frac{(2l+1)(2l-1)}{(l+m)(l-m)}
\right]^{1\over 2} \cos\theta \:_{s}Y_{l-1,m} \nonumber \\
&&-\left[\frac{(2l+1)(l+m-1)(l-m-1)(l-s-1)}{(2l-3)(l+m)(l-m)
    (l+s-1)}\right]^{1\over 2}\:_{s}Y_{l-2,m} \nonumber \\
&&+s\left[\frac{(2l+1)(2l-1)}{(l+m)(l-m)(l-s)(l+s-1)}\right]^{1\over 2}
    \sin\theta\:_{s-1}Y_{l-1,m}.
\label{recursion}
\end{eqnarray}
This will be used for evaluating the correlation functions in Sec.~\ref{result}.
Table~1 lists explicit expressions for some low-$l$ spin-weighted harmonics,
from which higher-$l$ ones can be constructed. 

The harmonics are related to the representation matrices of the 3-dimensional
rotation group. If we define a rotation $R(\alpha,\beta,\gamma)$ as being 
composed of a rotation $\alpha$ around $\hat e_3$, followed by $\beta$ around
the new $\hat e_2'$ and finally $\gamma$ around $\hat e_3''$,
the rotation matrix of $R$ will be given by~\cite{new}
\begin{equation}
D_{-sm}^{l}(\alpha,\beta,\gamma)=\sqrt{\frac{4\pi}{2l+1}}
\:_{s}Y_{lm}(\beta,\alpha)e^{-is\gamma}.
\end{equation}
Let us consider a rotation group multiplication, 
\begin{equation}
R(\alpha,\beta,-\gamma)= R(\phi',\theta',0) R^{-1}(\phi,\theta,0),
\end{equation}
where the angles are defined in Fig.~\ref{fig1}. 
In terms of rotation matrices, it becomes
\begin{equation}
D^l_{s_1 s_2}(\alpha,\beta,-\gamma)=\sum_{m} D^l_{s_1 m}
(\phi',\theta',0) D^{l*}_{s_{2} m}(\phi,\theta,0),
\end{equation}
which leads to the generalized addition theorem,\footnote
{This theorem was first derived in Eq.~(7) of Ref.~\cite{hu1}, which however
does not give correct signs for the geometric phase angles, $\alpha$ and
$\gamma$. Eq.~(\ref{addition}) will be useful in the following sections.}
\begin{equation}
\sum_{m}\:_{s_1}Y^*_{lm}(\theta',\phi')
          \:_{s_2}Y_{lm}(\theta,\phi) 
        =\sqrt{\frac{2l+1}{4\pi}}(-1)^{s_1-s_2}
           \:_{-s_1}Y_{ls_2}(\beta,\alpha)e^{-is_1\gamma}.
\label{addition}
\end{equation}

\section{Power Spectra}\label{power}
Following the notations in Ref.~\cite{sel1},
we expand the Stokes parameters as
\begin{eqnarray}
T(\hat n)&=&\sum_{lm}a_{T,lm}Y_{lm}(\hat n), \nonumber \\
Q(\hat n)-iU(\hat n)&=&\sum_{lm}a_{2,lm}\:_{2}Y_{lm}(\hat n), \nonumber \\
Q(\hat n)+iU(\hat n)&=&\sum_{lm}a_{-2,lm}\:_{-2}Y_{lm}(\hat n).
\label{expand}
\end{eqnarray}
The conjugation relation~(\ref{conjugation}) requires that
\begin{equation}
a_{T,lm}^*=(-1)^m a_{T,l-m},\quad a_{-2,lm}^*=(-1)^m a_{2,l-m}.
\end{equation}
For Stokes parameters in CMB measurements, using the parity 
relation~(\ref{parity}), we have
\begin{eqnarray}
{\cal T}(\hat x)&=&\sum_{lm}(-1)^l a_{T,lm}Y_{lm}(\hat x), \nonumber \\
{\cal Q}(\hat x)+i{\cal U}(\hat x)&=&\sum_{lm}(-1)^l a_{2,lm}
                                     \:_{-2}Y_{lm}(\hat x), \nonumber \\
{\cal Q}(\hat x)-i{\cal U}(\hat x)&=&\sum_{lm}(-1)^l a_{-2,lm}
                                     \:_{2}Y_{lm}(\hat x).
\end{eqnarray}
Isotropy in the mean guarantees the following ensemble averages:
\begin{eqnarray}
\left<a^{*}_{T,l'm'}a_{T,lm}\right>&=&C_{Tl}\delta_{l'l}\delta_{m'm},
\nonumber \\
\left<a^{*}_{2,l'm'}a_{2,lm}\right>&=&(C_{El}+C_{Bl})\delta_{l'l}
\delta_{m'm}, \nonumber \\ 
\left<a^{*}_{2,l'm'}a_{-2,lm}\right>&=&(C_{El}-C_{Bl})
\delta_{l'l}\delta_{m'm}, \nonumber \\
\left<a^{*}_{T,l'm'}a_{2,lm}\right>&=&-C_{Cl}\delta_{l'l}\delta_{m'm}.
\label{CMBaa}
\end{eqnarray}
Consider two points $\hat n'(\theta',\phi')$ and 
$\hat n(\theta,\phi)$ on the sphere.  
Using the addition theorem~(\ref{addition}) and Eq.~(\ref{CMBaa}), 
we obtain the correlation functions,
\begin{eqnarray}
&&\left<T^{*}(\hat n')T(\hat n)\right>
  =\sum_l\frac{2l+1}{4\pi}C_{Tl} P_l(\cos\beta), \label{ct} \\
&&\left<T^{*}(\hat n')[Q(\hat n)+iU(\hat n)]\right>
  =-\sum_l\frac{2l+1}{4\pi}\sqrt{\frac{(l-2)!}{(l+2)!}}C_{Cl}
   P^2_l(\cos\beta) e^{2i\alpha}, \label{cc}\\
&&\left<[Q(\hat n')+iU(\hat n')]^*[Q(\hat n)+iU(\hat n)]\right>
  =\sum_{l}\sqrt{\frac{2l+1}{4\pi}}(C_{El}+C_{Bl})
   \:_{2}Y_{l-2}(\beta,0)e^{2i(\alpha-\gamma)}, \label{c+} \\
&&\left<[Q(\hat n')-iU(\hat n')]^*[Q(\hat n)+iU(\hat n)]\right>
  =\sum_{l}\sqrt{\frac{2l+1}{4\pi}}(C_{El}-C_{Bl})
   \:_{2}Y_{l2}(\beta,0)e^{2i(\alpha+\gamma)}, \label{c-}
\end{eqnarray}
where $\alpha$, $\beta$, and $\gamma$ are the angles defined in 
Fig.~\ref{fig1}.
Eq.~(\ref{cc}) is the most general form of those found in 
Refs.~\cite{cri1,ng4,sel,mel}. 
In the small-angle approximation, i.e. $\beta<<1$,
$\alpha\simeq\gamma$, so Eq.~(\ref{c+}) depends only on the separation 
angle $\beta$. When $\hat n'$ and $\hat n$ lie on the same longitude,
$\alpha=\gamma=0$ and hence Eqs.~(\ref{cc},\ref{c+},\ref{c-}) depend 
only on $\beta$. When $\hat n'$ and $\hat n$ lie on the same latitude,
$\alpha+\gamma=\pi$. Hence the phase angle in Eq.~(\ref{c-}) vanishes, and
that in Eq.~(\ref{c+}) becomes equal to $e^{4i\alpha}$. 

A coordinate-independent set
of correlation functions has been obtained by defining correlation functions
of Stokes parameters $(Q_r,U_r)$ with respect to axes which are parallel and 
perpendicular to the great arc connecting the two points being 
correlated~\cite{kam2}. 
This prescription is indeed equivalent to the transformations:
\begin{eqnarray}
Q_{r}(\hat n')+iU_{r}(\hat n')&=&
              e^{-2i\gamma}[Q(\hat n')+iU(\hat n')],\nonumber\\
Q_{r}(\hat n)+iU_{r}(\hat n)&=&e^{-2i\alpha}[Q(\hat n)+iU(\hat n)].
\end{eqnarray}
The authors in Ref.~\cite{kam2} expanded $Q_r$ and $U_r$ in terms of
tensor spherical harmonics.
To calculate the two-point correlation functions between 
$T$, $Q_r$, and $U_r$, they chose
one point to be at the north pole and the other on the $\phi=0$
longitude, and argued that the correlation functions depend only on 
the angular separation of the two points. Then, they had to evaluate the 
asymptotic forms for the tensor spherical harmonics at the north pole.
Their results are simply equal to the above correlation functions 
without the phase angles. Here, using the compact generalized addition
theorem~(\ref{addition}), 
we have given a general and straightforward way of obtaining the
correlation functions. In addition, the phase information is retained. 
We will see in Sec.~\ref{estimator} and Sec.~\ref{like} that these phase
angles can be easily removed or evaluated in taking experimental data.   
 
Therefore, the statistics of the CMB anisotropy and polarization is fully
described by four independent power spectra $(C_{Tl}, C_{El}, C_{Bl}, C_{Cl})$
or their corresponding correlation functions. Here, we outline how to evaluate
the spectra. The details can be found in Refs.~\cite{sel2,kam2}.

Since the four spectra are rotationally invariant, it suffices to consider 
the contribution from a single $\hat k$-mode of the perturbation, 
and then integrate over all the modes. In particular, the calculation 
will be greatly simplified if we choose $\hat k=\hat e_3$. 
For scalar perturbations, the contribution of the
$\hat k$-mode to $(T(\hat n),Q(\hat n),U(\hat n))$ is
$(\Delta^{(S)}_T,\Delta^{(S)}_P,0)$. 
For tensor perturbations, the contribution is
\begin{equation}
\left(\begin{array}{c} (1-\cos^2\theta)\cos2\phi\,\Delta^{(T)}_T\\
                       (1+\cos^2\theta)\cos2\phi\,\Delta^{(T)}_P\\
                       2\cos\theta\sin2\phi\,\Delta^{(T)}_P
      \end{array} \right),
\end{equation}
for $+$-mode. The $\times$-mode contribution is from making the replacements,
$\cos2\phi\to \sin2\phi$ and $\sin2\phi\to -\cos2\phi$. The quantities
$\Delta$'s are then computed by solving the Boltzmann hierarchy equations 
or by the line-of-sight integration method~\cite{zal2}. 
In the following sections, we will use CMBFAST Boltzmann code~\cite{zal2} 
to evaluate all $C_{Xl}$'s.

\section{Window Function}\label{window}

Due to the finite beam size of the antenna, any information on angular
scales less than about the beam width is smeared out. This effect can be
approximated by a Gaussian response function,
\begin{equation}
dR(\beta,\alpha)=\frac{\beta\,d\beta\,d\alpha}{2\pi\sigma_b^2}\;
e^{-\frac{\beta^2}{2\sigma_b^2}},
\end{equation}
where $\sigma_b$, much less than $1$, 
is the Gaussian beam width of the antenna,
$\beta$ and $\alpha$ are spherical polar angles with respect to a polar axis 
along the direction $\hat n(\theta,\phi)$. 
Therefore, a measurement can be represented
as a convolution of the response function and the expected Stokes parameters,
\begin{equation}
\int dR(\beta,\alpha) X(\theta',\phi'),
\end{equation}
where $X$ denotes $T$, or $Q\pm iU$. This can be accounted by a 
mapping of the harmonics in Eq.~(\ref{expand}),
\begin{equation}
\:_{s}Y_{lm}(\theta,\phi) \to 
\int dR(\beta,\alpha) \:_{s}Y_{lm}(\theta',\phi').
\label{mapping}
\end{equation}
From Eq.~(\ref{addition}), we have
\begin{equation}
\:_{s}Y_{lm}(\theta',\phi')=\sqrt{\frac{4\pi}{2l+1}}
\sum_{m'}\:_{s}Y_{lm'}(\beta,\alpha)\,e^{is\gamma}\:_{-m'}Y_{lm}(\theta,\phi).
\end{equation}
Therefore, the convolution involves the integral,
\begin{equation}
\sqrt{\frac{4\pi}{2l+1}} \int dR(\beta,\alpha)
\:_{s}Y_{lm'}(\beta,\alpha)\,e^{is\gamma}.
\label{integral}
\end{equation}
Making the approximation that $\alpha\simeq \gamma$ for $\sigma_b<<1$ and
using the explicit expression~(\ref{s-har}), 
the integral~(\ref{integral}) has a series solution as
\begin{eqnarray}
&&(-1)^s \left\{1-\left[(l-s)(l+s)+l\right]\left(\frac{\sigma_b^2}{2}\right)
+\left[\frac{1}{2}(l-s)(l-s-1)(l+s)(l+s-1)\right.\right. \nonumber \\
&&\left.\left.-(l-s)(l+s)\left(-2l+\frac{4}{3}\right)
+2\left(-\frac{l}{6}+\frac{l^{2}}{2}\right)\right] 
\left(\frac{\sigma_b^2}{2}\right)^{2}+\;...\right\} \delta_{-m',s}\nonumber\\
&\simeq&\; (-1)^s
\exp\left[-\left(l(l+1)-s^2\right)\frac{\sigma_b^2}{2}\right]\;\delta_{-m',s}.
\end{eqnarray}
Hence, the mapping~(\ref{mapping}) is approximated by
\begin{equation}
\:_{s}Y_{lm}(\theta,\phi)\to (-1)^s \:_{s}W_l^{1\over 2} 
\:_{s}Y_{lm}(\theta,\phi),
\end{equation}
where $\:_{s}W_l$ is the window function,
\begin{equation}
\:_{s}W_l= \exp\left[-\left(l(l+1)-s^2\right)\sigma_b^2\right].
\label{swl}
\end{equation}
When $s=0$, it reduces to the usual window function in anisotropy case, 
\begin{equation} 
\:_{0}W_l\equiv W_l=\exp[-l(l+1)\sigma_b^2].
\end{equation} 
The approximation $\:_{s}W_l\simeq \exp[-l^2\sigma_b^2]$ works very well 
for high $l$'s.

\section{Instrumental Noise}\label{noise}

In the CMB experiment, a pixelized map of the 
CMB smoothed with a Gaussian beam
is created. In each pixel, the signal has a 
contribution from the CMB and from 
the instrumental noise. A convenient way of describing the amount of
instrumental noise is to specify the rms noise per pixel $\sigma_{\rm pix}$, 
which depends on the detector
sensitivity $s$ and the time spent observing each pixel $t_{\rm pix}$:
$\sigma_{\rm pix}=s/\sqrt{t_{\rm pix}}$. The noise in each pixel is 
uncorrelated with that in any other pixel, and is uncorrelated 
with the CMB component.
Let $\Omega_{\rm pix}$ be the solid angle subtended by a pixel. 
Usually, given a total observing time, $t_{\rm pix}$
is directly proportional to $\Omega_{\rm pix}$. Thus, we can define a quantity
$w^{-1}$, the inverse statistical weights per unit solid angle, to measure
the experimental sensitivity independent of pixel size~\cite{kno}:
\begin{equation}
w^{-1}=\Omega_{\rm pix} \sigma_{\rm pix}^2.
\end{equation}

Let us simulate the instrumental noise with a background of white noise
superposed upon the microwave sky. The statistics of the white noise is
completely determined by
\begin{eqnarray}
\left<a^{N\;*}_{T,l'm'}a^N_{T,lm}\right> 
       &=&w_T^{-1}\delta_{l'l}\delta_{m'm}, \nonumber \\
\left<a^{N\;*}_{2,l'm'}a^N_{2,lm}\right> 
       &=&2w_P^{-1}\delta_{l'l}\delta_{m'm}, \nonumber \\
\left<a^{N\;*}_{-2,l'm'}a^N_{-2,lm}\right> 
       &=&2w_P^{-1}\delta_{l'l}\delta_{m'm}, \nonumber \\
\left<a^{N\;*}_{T,l'm'}a^N_{\pm2,lm}\right> 
&=&\left<a^{N\;*}_{2,l'm'}a^N_{-2,lm}\right>=0,
\label{Naa}
\end{eqnarray}
where the label $N$ stands for noise, $w_T^{-1}$ and $w_P^{-1}$ are 
constants to be dertermined. Then, the two-point correlation functions are
\begin{eqnarray}
&&\left<T^N(\hat n')T^N(\hat n)\right>
  =\sum_l\frac{2l+1}{4\pi} w_T^{-1} W_l P_l(\cos\beta), \nonumber \\
&&\left<[Q^N(\hat n')+iU^N(\hat n')]^*[Q^N(\hat n)+iU^N(\hat n)]\right>
  =\sum_{l}\sqrt{\frac{2l+1}{4\pi}} 2w_P^{-1} \:_{2}W_l
   \:_{2}Y_{l-2}(\beta,0)e^{2i(\alpha-\gamma)}, \nonumber \\
&&\left<[Q^N(\hat n')-iU^N(\hat n')]^*[Q^N(\hat n)+iU^N(\hat n)]\right> =0.
\label{Ncf}
\end{eqnarray}
Defining $\sigma^T$ and $\sigma^P$ be the rms anisotropy and polarization
variances respectively, 
for small beam width we obtain from Eqs.~(\ref{swl},\ref{Ncf}) that 
\begin{eqnarray}
&&(\sigma^T)^2\equiv \left<{T^N}^2\right>=\frac{w_T^{-1}}{4\pi\sigma_b^2},
      \nonumber \\
&&(\sigma^P)^2\equiv \left<{Q^N}^2\right>=\left<{U^N}^2\right>=
      \frac{w_P^{-1}}{4\pi\sigma_b^2}.
\end{eqnarray}
Therefore, if we assume $\Omega_{\rm pix}=4\pi\sigma_b^2$, then 
the variances would be the pixel noise, and $w_{T,P}^{-1}$ be the inverse
statisical weights per unit soild angle:
\begin{equation}
\sigma^{T,P}=\sigma^{T,P}_{\rm pix},\quad 
w_{T,P}^{-1}=\Omega_{\rm pix}\left(\sigma^{T,P}_{\rm pix}\right)^2.
\end{equation}

If both anisotropy and polarization are obtained from the same experiment by
adding and subtracting the two orthogonal linear polarization states given 
equal integration times, then
\begin{equation} 
\left(\sigma^T_{\rm pix}\right)^2={1\over 2}\left(\sigma^P_{\rm pix}\right)^2.
\end{equation}
If they are from different maps, the noise is uncorrelated.

\section{Full-sky Correlation Function Estimators}\label{estimator}

The CMB map is inevitably contaminated by instrumental noise and
other known or unresolved foreground sources. However, the foreground 
contamination can be removed by observing the CMB at multi-frequencies and
detecting its unique spectral dependence. After the removal of background
contamination, the microwave map (denoted by label $M$) 
is made of the genuine CMB and  
instrumental noise:
\begin{equation}
a^M_{T,lm}=a_{T,lm}+a^N_{T,lm},\quad 
a^M_{\pm 2,lm}=a_{\pm 2,lm}+a^N_{\pm 2,lm}.
\end{equation}
Thus the statistics of the noisy CMB map is induced from that of the CMB 
in Eq.~(\ref{CMBaa}) and that of the noise in Eq.~(\ref{Naa}). 
Again note that the noise is uncorrelated with the CMB signal, i.e. 
$\left<a^N a\right>=0$.

Now we are going to construct the full-sky averaged correlation function
estimators. Let us begin taking an average of a product of two spherical
harmonics over the whole sky, 
\begin{eqnarray}
\left\{Y^*_{l'm'}(\hat n')Y_{lm}(\hat n)\right\}_S&\equiv&
\int d\Omega' d\Omega \,Y^*_{l'm'}(\hat n')Y_{lm}(\hat n) \nonumber \\
&=& \frac{1}{4\pi} P_l(\cos\beta) \delta_{l'l} \delta_{m'm},
\label{YY}
\end{eqnarray}
where the curly brackets $\{\}_S$ denote a full-sky averaging at a fixed 
separation angle $\beta$. The sky averaging can be done easily using 
Eqs.~(\ref{addition},\ref{ortho}). We firstly transform $Y_{l'm'}(\hat n')$
defined by a spherical coordinate system $\hat e_3$ to a new
coordinate system $\hat n$, and then performing azimuthal integration by
rotating the transformed $\hat n'$ about $\hat n$ with a fixed separation
angle $\beta$. Finally, the remaining product of spherical harmonics with
angle variables $\hat n$ is integrated over the whole sky.

To generalize the averaging procedure to spin-$s$ spherical harmonics,
some complications have to be taken. As we have seen in 
Eqs.~(\ref{cc},\ref{c+},\ref{c-}), multiplication of higher spin 
harmonics depend explicitly on local angles. 
Therefore, we define the full-sky averaging as
\begin{eqnarray}
\left\{\:_{s_1}Y^*_{l'm'}(\hat n')\:_{s_2}Y_{lm}(\hat n)\right\}_S&\equiv&
\int d\Omega' d\Omega \:_{s_1}Y^*_{l'm'}(\hat n')\:_{s_2}Y_{lm}(\hat n) 
e^{i(s_1\gamma - s_2\alpha)} \nonumber \\
&=&\sqrt{\frac{1}{4\pi(2l+1)}}\:_{s_1}Y_{l-s_2}(\beta,0)
\delta_{l'l}\delta_{m'm}.
\label{sky}
\end{eqnarray}
Obviously, when $s_1=s_2=0$, it reduces to Eq.~(\ref{YY}). 

We define four full-sky averaged correlation function estimators,
\begin{eqnarray}
{\cal C}_T(\beta)&\equiv& \left\{T^{M*}(\hat n') T^M(\hat n)\right\}_S
   -\left<\left\{T^{N*}(\hat n') T^N(\hat n)\right\}_S\right> \nonumber \\
  &=&\sum_l\frac{2l+1}{4\pi}\left({\cal C}^M_{Tl} -w_T^{-1}\right)
       W_l P_l(\cos\beta), \nonumber \\
{\cal C}_C(\beta)&\equiv& {1\over 2}\left\{T^{M*}(\hat n')
      [Q^M(\hat n)+iU^M(\hat n)]+{\rm h.c.} \right\}_S \nonumber \\
 &=&-\sum_l\frac{2l+1}{4\pi}\sqrt{\frac{(l-2)!}{(l+2)!}}
   {1\over 2}\left({\cal C}^M_{Cl}+{\cal C}^{M*}_{Cl}\right)
  W_l^{1\over 2} \:_{2}W_l^{1\over 2} P^2_l(\cos\beta), \nonumber \\
{\cal C}_+(\beta)&\equiv& \left\{[Q^M(\hat n')+iU^M(\hat n')]^*
     [Q^M(\hat n)+iU^M(\hat n)]\right\}_S -
     \left<\left\{[Q^N(\hat n')+iU^N(\hat n')]^*
     [Q^N(\hat n)+iU^N(\hat n)]\right\}_S\right> \nonumber \\
 &=&\sum_{l}\sqrt{\frac{2l+1}{4\pi}}\left({\cal C}^M_{+l}
     -2w_P^{-1}\right) \:_{2}W_l \:_{2}Y_{l-2}(\beta,0), \nonumber \\
{\cal C}_-(\beta)&\equiv& {1\over 2}\left\{[Q^M(\hat n')-
      iU^M(\hat n')]^* [Q^M(\hat n)+iU^M(\hat n)]
      + {\rm h.c.} \right\}_S \nonumber \\
 &=&\sum_{l}\sqrt{\frac{2l+1}{4\pi}}{1\over 2}\left({\cal C}^M_{-l}
     +{\cal C}^{M*}_{-l}\right) \:_{2}W_l \:_{2}Y_{l2}(\beta,0),
\label{calC}
\end{eqnarray}
where
\begin{eqnarray}
{\cal C}^M_{Tl}&\equiv&\frac{1}{2l+1}\sum_{m}a^{M*}_{T,lm}a^M_{T,lm},
 \nonumber \\
{\cal C}^M_{Cl}&\equiv&-\frac{1}{2l+1}\sum_{m}a^{M*}_{T,lm}a^M_{2,lm},
\nonumber \\
{\cal C}^M_{\pm l}&\equiv&\frac{1}{2l+1}\sum_{m}a^{M*}_{\pm 2,lm}a^M_{2,lm}.
\end{eqnarray}
The ensemble mean of each estimator is 
\begin{eqnarray}
\left<{\cal C}_T(\beta)\right>&=&
   \sum_l\frac{2l+1}{4\pi}C_{Tl} W_l P_l(\cos\beta), \nonumber\\
\left<{\cal C}_C(\beta)\right>&=&
   -\sum_l\frac{2l+1}{4\pi}\sqrt{\frac{(l-2)!}{(l+2)!}}C_{Cl}
     W_l^{1\over 2} \:_{2}W_l^{1\over 2} P^2_l(\cos\beta),\nonumber \\
\left<{\cal C}_{\pm}(\beta)\right>&=&
   \sum_{l}\sqrt{\frac{2l+1}{4\pi}}(C_{El}\pm C_{Bl})
   \:_{2}W_l \:_{2}Y_{l\mp2}(\beta,0).
\label{mean}
\end{eqnarray}
And the covariance matrix can be constructed as  
\begin{equation}
{\bf M}_{X'Y}\equiv
\left<[{\cal C}_X(\beta')-\left<{\cal C}_X(\beta')\right>]
[{\cal C}_Y(\beta) - \left<{\cal C}_Y(\beta)\right>]\right>,
\label{cm}
\end{equation}
where $X,Y=T,C,+,-$. Here the prime denotes a different separation 
angle. The diagonal entries are given by
\begin{eqnarray}
{\bf M}_{T'T}&=&
  \frac{1}{8\pi^2}\sum_l (2l+1)\left(C_{Tl}+w_T^{-1}\right)^2 
  W_l^2 P_l(\cos\beta') P_l(\cos\beta), \nonumber \\
{\bf M}_{C'C}&=&
  \frac{1}{4\pi}\sum_l \frac{2l+1}{4\pi} \frac{(l-2)!}{(l+2)!}
  \left[C_{Cl}^2+\left(C_{Tl}+w_T^{-1}\right)
  \left(C_{El}+w_P^{-1}\right)\right]
  W_l \:_{2}W_l P^2_l(\cos\beta') P^2_l(\cos\beta),\nonumber \\
{\bf M}_{+'+}&=&
  \frac{1}{2\pi}\sum_l \left[C_{El}^2+C_{Bl}^2+2 \left(w_P^{-1}\right)^2
  + 2\left(C_{El}+C_{Bl}\right) w_P^{-1} \right]
  \:_{2}W_l^2 \:_{2}Y_{l-2}(\beta',0) \:_{2}Y_{l-2}(\beta,0),\nonumber\\
{\bf M}_{-'-}&=&
  \frac{1}{2\pi}\sum_l \left[C_{El}^2+C_{Bl}^2+2 \left(w_P^{-1}\right)^2
  + 2\left(C_{El}+C_{Bl}\right) w_P^{-1} \right]
  \:_{2}W_l^2 \:_{2}Y_{l2}(\beta',0) \:_{2}Y_{l2}(\beta,0).
\label{variance}
\end{eqnarray}
The off-diagonal entries are similarly calculated. In particular, the 
off-diagonal term of the submatrix ${\bf M}_{X'Y}$~($X,Y= +,-$) is 
\begin{equation}
{\bf M}_{+'-}=
  \frac{1}{2\pi}\sum_l \left[C_{El}^2-C_{Bl}^2+ 2\left(C_{El}-C_{Bl}\right)
   w_P^{-1} \right]
  \:_{2}W_l^2 \:_{2}Y_{l-2}(\beta',0) \:_{2}Y_{l2}(\beta,0).
\end{equation}

In practical situations, 
a galaxy-cut on the CMB map is necessary due to radiation
pollution along the galactic plane, and due to limited observation time, 
usually only a fraction of the sky would be sampled. 
For instance, the effective 
CMB coverage of the {\it COBE} DMR is $4\pi f$, where $f\simeq 2/3$. 
This incomplete sky coverage would generally induce a sample variance,
whose size depends both on the experimental sampling strategy and the 
underlying power spectra of the fluctuations. It was found that
the covariances calculated above scale roughly
with sky coverage as $f^{-1}$ for small-scale experiments~\cite{sco}. 
For large-scale experiments such as the {\it COBE} DMR, they scale roughly as
\begin{equation}
0.446+0.542f^{-1}-0.0079f^{-2},
\end{equation}
valid for $f^{-1}<15$~\cite{ng}. 
The difference from $f^{-1}$ scaling is mainly due a large correlation
angle in the large-scale experiment.

\section{Correlation Measurements in Future Missions}\label{result}

The MAP and Planck missions plan to measure all-sky CMB anisotropy
and polarization. It has been discussed how to construct the optimal 
estimators for the power spectra corrected for 
noise bias, and their corresponding variances from the all-sky 
map~\cite{sel2,kam2,hu2}. 
To estimate the level of signal and noise, 
we hereby give an alternative real-space analysis,
evaluating the ensemble means and variances of the full-sky averaged
correlation function estimators for the MAP and Planck configurations, i.e.
\begin{equation}
C_X(\theta)\equiv\left<{\cal C}_X\right>,\quad
\Delta C_X (\theta)\equiv {\bf M}_{XX}^{1\over 2},
\end{equation}
which are respectively given by Eq.~(\ref{mean}), and Eq.~(\ref{variance})
with $\theta'=\theta$.

\begin{figure}
\leavevmode
\hbox{
\epsfxsize=6.5in
\epsffile{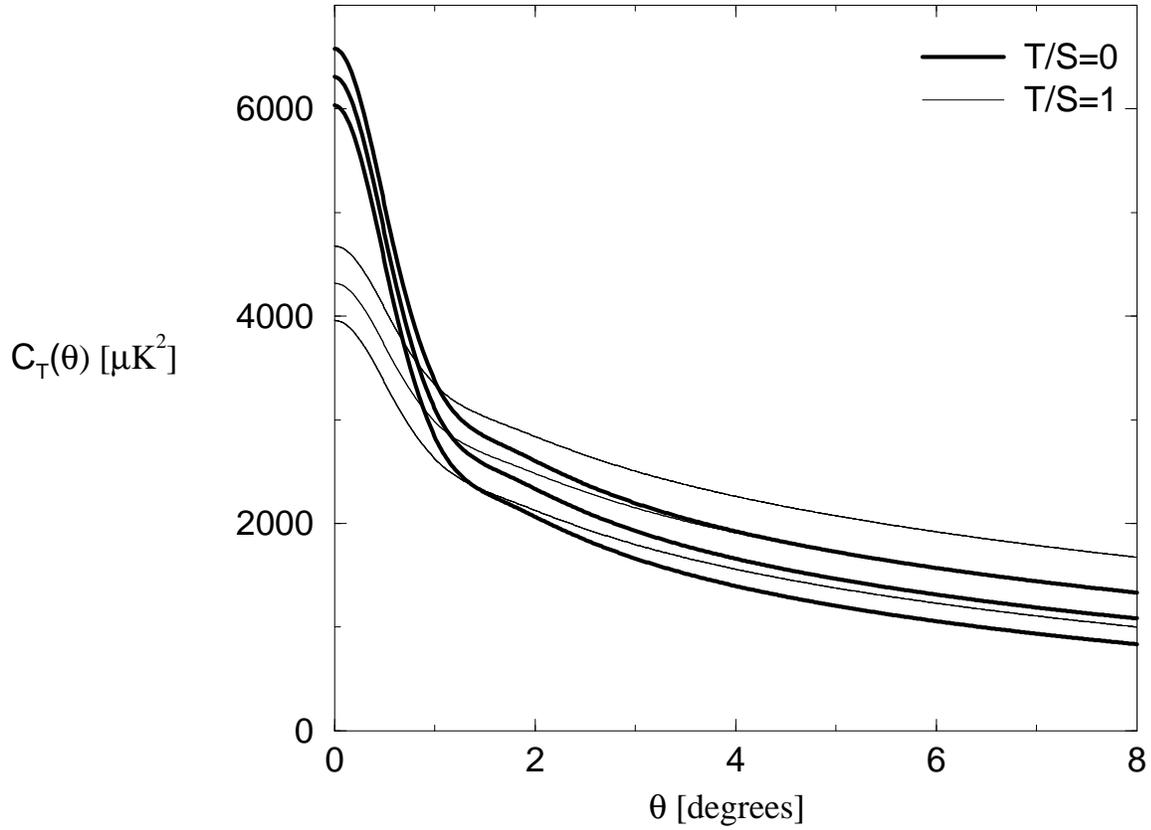}}
\caption{Full-sky averaged correlation function and its variance, 
$C_T(\theta)\pm \Delta C_T (\theta)$, versus separation angle $\theta$
for MAP and Planck experiments.
In each model, the correlation function is denoted by a middle line 
sandwiched by a pair of $1 \sigma$ lines, which represent the theoretical
error expected in MAP and Planck experiments. This means that the error
is dominated by cosmic variance rather than instrumental noise.}
\label{fig2}
\end{figure}

\begin{figure}
\leavevmode
\hbox{
\epsfxsize=6.5in
\epsffile{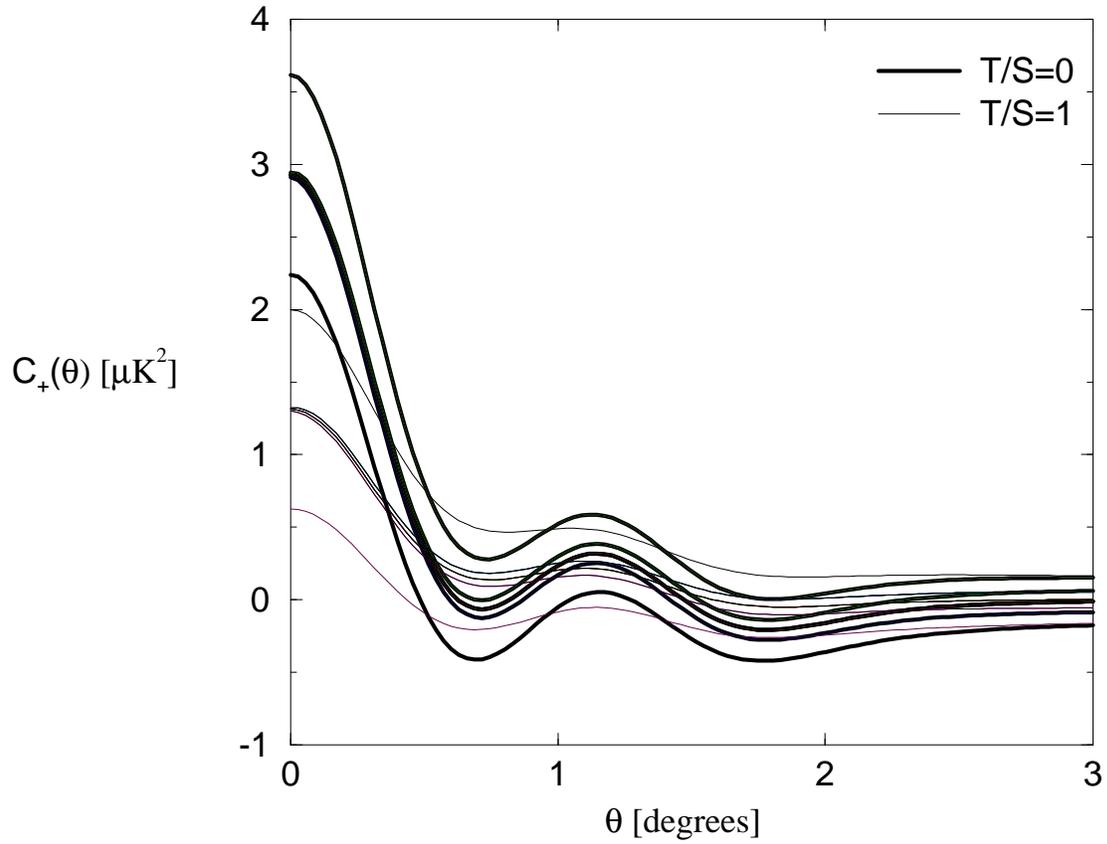}}
\caption{Full-sky averaged correlation function 
$C_+(\theta)\pm \Delta C_+ (\theta)$.
In each model, the correlation function is denoted by a middle line 
sandwiched by two pairs of $1 \sigma$ lines. The outer pair of lines is 
for the MAP configuration, while the inner pair for the Planck Surveyor.} 
\label{fig3}
\end{figure}

\begin{figure}
\leavevmode
\hbox{
\epsfxsize=6.5in
\epsffile{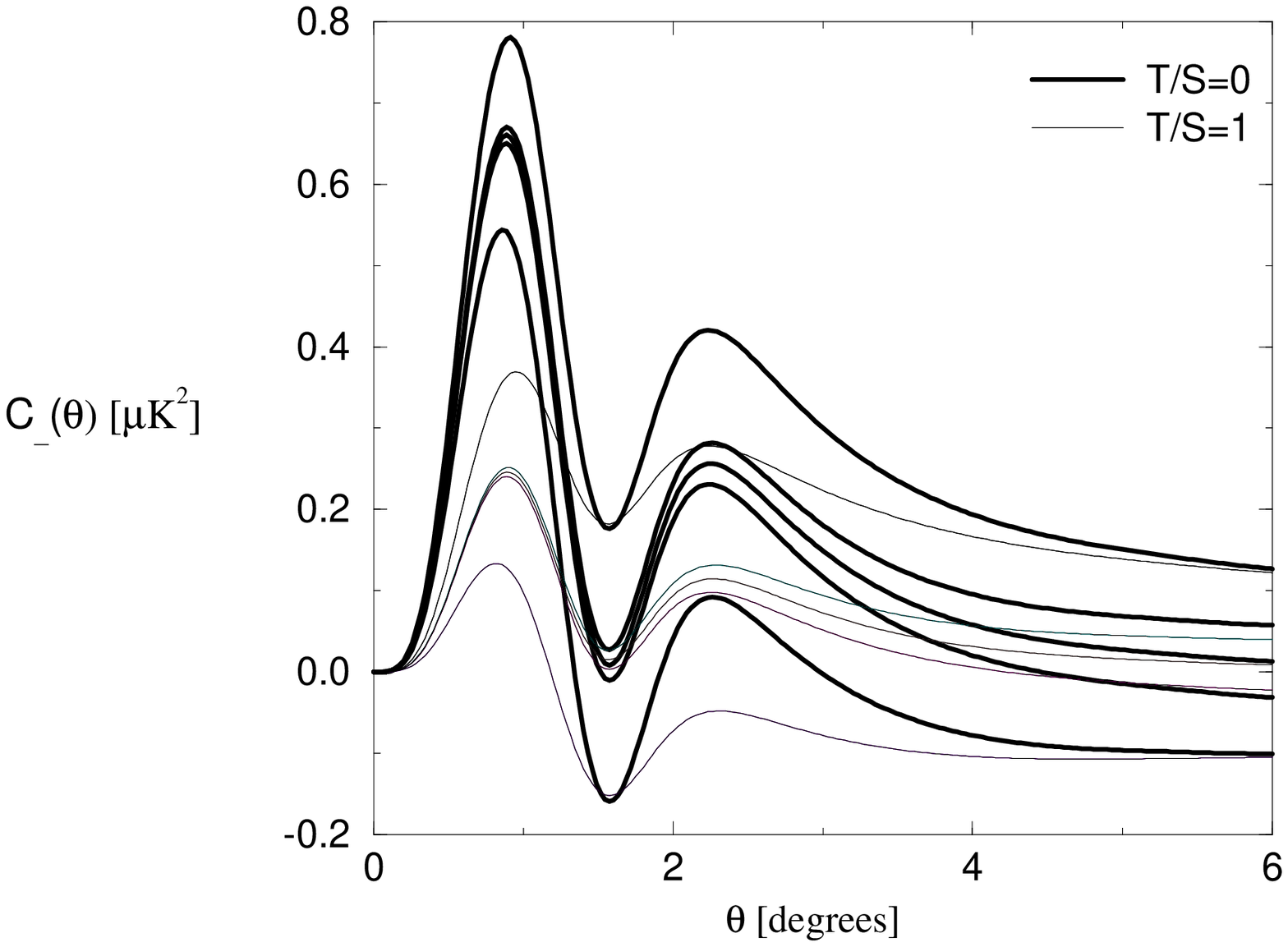}}
\caption{Full-sky averaged correlation function 
$C_-(\theta)\pm \Delta C_- (\theta)$ for MAP and Planck.}
\label{fig4}
\end{figure}

\begin{figure}
\leavevmode
\hbox{
\epsfxsize=6.5in
\epsffile{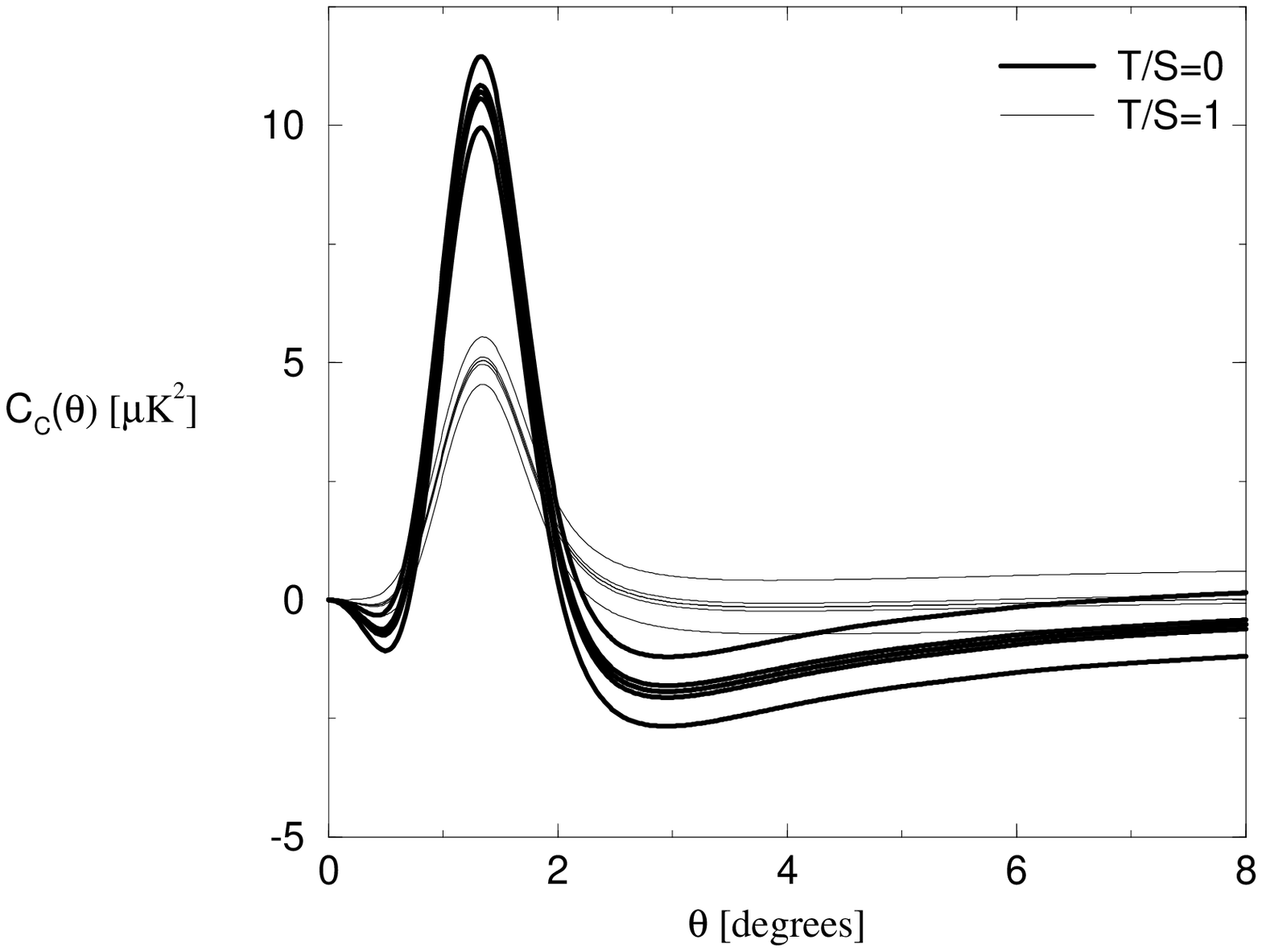}}
\caption{Full-sky averaged correlation function 
$C_C(\theta)\pm \Delta C_C (\theta)$ for MAP and Planck.}
\label{fig5}
\end{figure}

We assume the standard cold dark matter (sCDM) model: $\Omega_0=1$, $h=0.5$,
$\Omega_B h^2=0.0125$, and no reionization after the hydrogen recombination.
Two extreme cases are evaluated: $T/S=0$ and $T/S$=1,
where $T$ and $S$ are the anisotropy quadrupole moments induced respectively
by tensor and scalar perturbations. All power spectra are computed by the 
CMBFAST code. The recursion relation~(\ref{recursion}) has been used 
for evaluating the spherical harmonics
in the correlation functions. The results are plotted in 
Figs.~\ref{fig2}-\ref{fig5}, which are respectively $C_X(\theta)$ attached
with its variance $\Delta C_X (\theta)$, where $X=T,+,-,C$.
In making the plots, we have used beam width $\theta_{FWHM}=0.5^o$, where
$\sigma_b=0.425\times \theta_{FWHM}$.
Typical values of the experimental sensitivity for MAP are 
$w_{T}^{-1}=(0.1{\mu K})^2$ and $w_{P}^{-1}=(0.15{\mu K})^2$, 
while for Planck they are about a factor of 100 smaller.
In Figs.~\ref{fig2}-\ref{fig5}, the thick and thin solid lines 
represent the cases with 
$T/S=0$ and $T/S=1$ respectively. In each case, 
the ensemble average is denoted by a middle line sandwiched by 
two pairs of $1 \sigma$ lines. 
The outer pair of lines is for the MAP satellite 
while the inner pair for the Planck Surveyor. Note that in Fig.~\ref{fig2}
the two pairs of $1 \sigma$ lines merge into a single pair, 
which means that the noise is dominated by cosmic variance rather than 
instrumental noise.
  
The theoretical expectation of the rms polarization signal in sCDM models, 
$[C_+(0)]^{1/2}$, is at a level of $1 \mu K$. For the MAP experiment,
the polarization signal to noise ratio S/N is about $1-2$.
The S/N ratio of the anisotropy-polarization correlation $C_C(\theta)$
is about $3-4$ at $\theta\simeq 1.3^o$, and the absence of tensor mode
makes the cross-correlation significantly negative on few-degree scales.
For Planck the corresponding S/N ratios are much higher. The MAP would likely
detect the anisotropy-polarization correlation, which however is not sensitive
to $C_{El}$ or $C_{Bl}$. The detection of the electric and magnetic
components would require the Planck satellite.

\begin{figure}
\leavevmode
\hbox{
\epsfxsize=6.5in
\epsffile{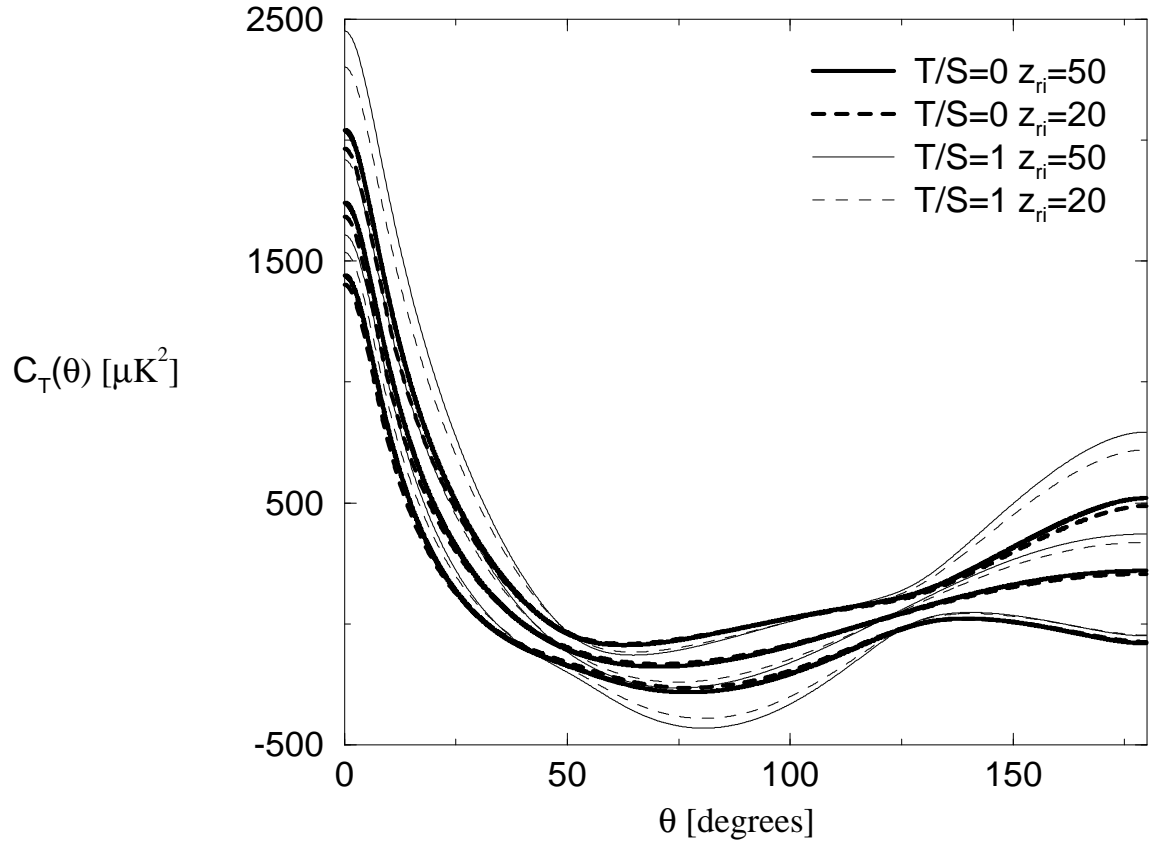}}
\caption{Full-sky averaged correlation function and its variance, 
$C_T(\theta)\pm \Delta C_T (\theta)$, for SPOrt/ISS experiment.
In each model, the correlation function is denoted by a middle line 
sandwiched by a pair of $1 \sigma$ lines. Solid and dashed lines denote 
respectively the models with reionization redshifts $z_{\rm ri}=50$ and $20$.}
\label{fig6}
\end{figure}

\begin{figure}
\leavevmode
\hbox{
\epsfxsize=6.5in
\epsffile{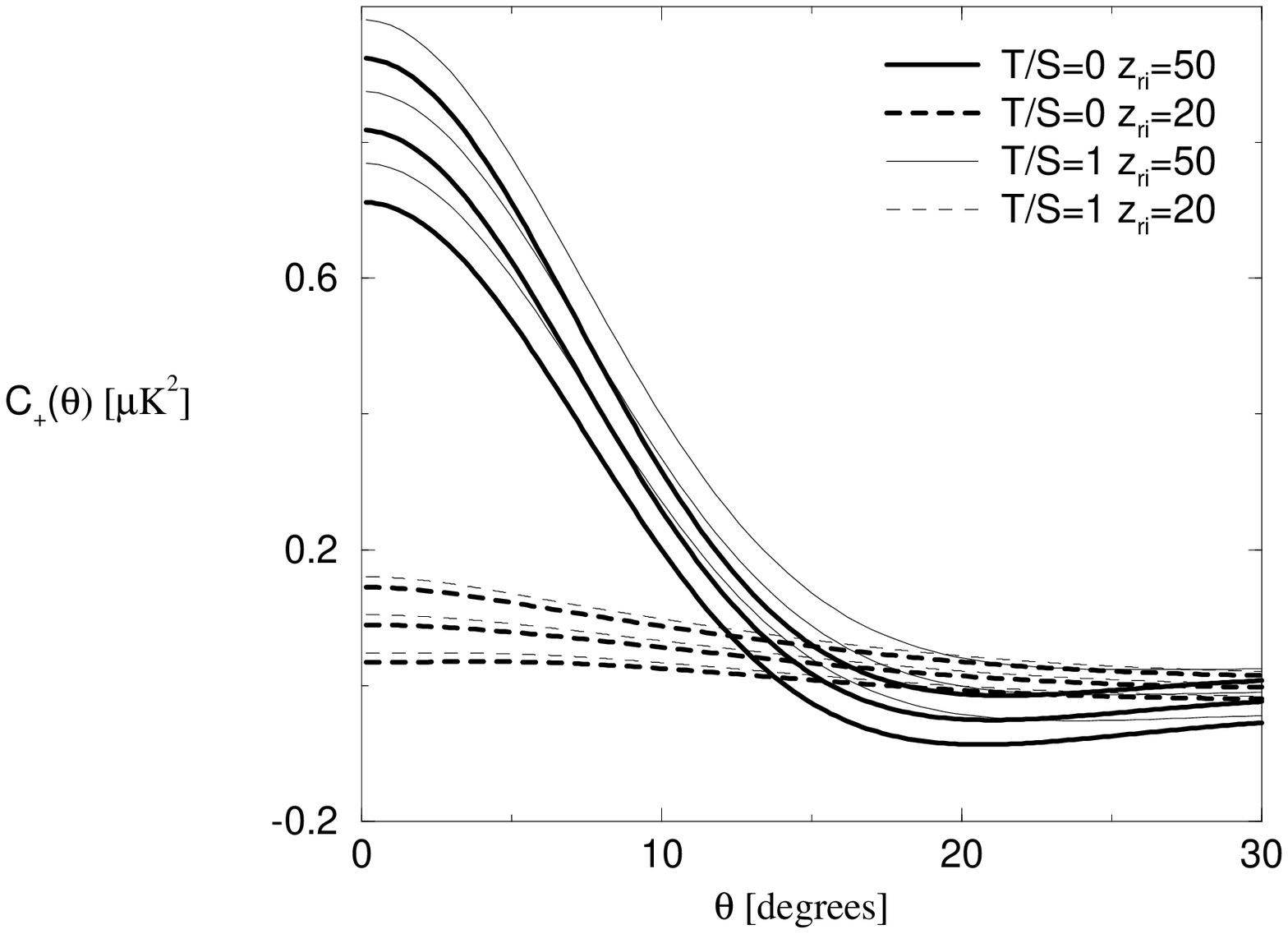}}
\caption{Full-sky averaged correlation function 
$C_+(\theta)\pm \Delta C_+ (\theta)$ for SPOrt/ISS.} 
\label{fig7}
\end{figure}

\begin{figure}
\leavevmode
\hbox{
\epsfxsize=6.5in
\epsffile{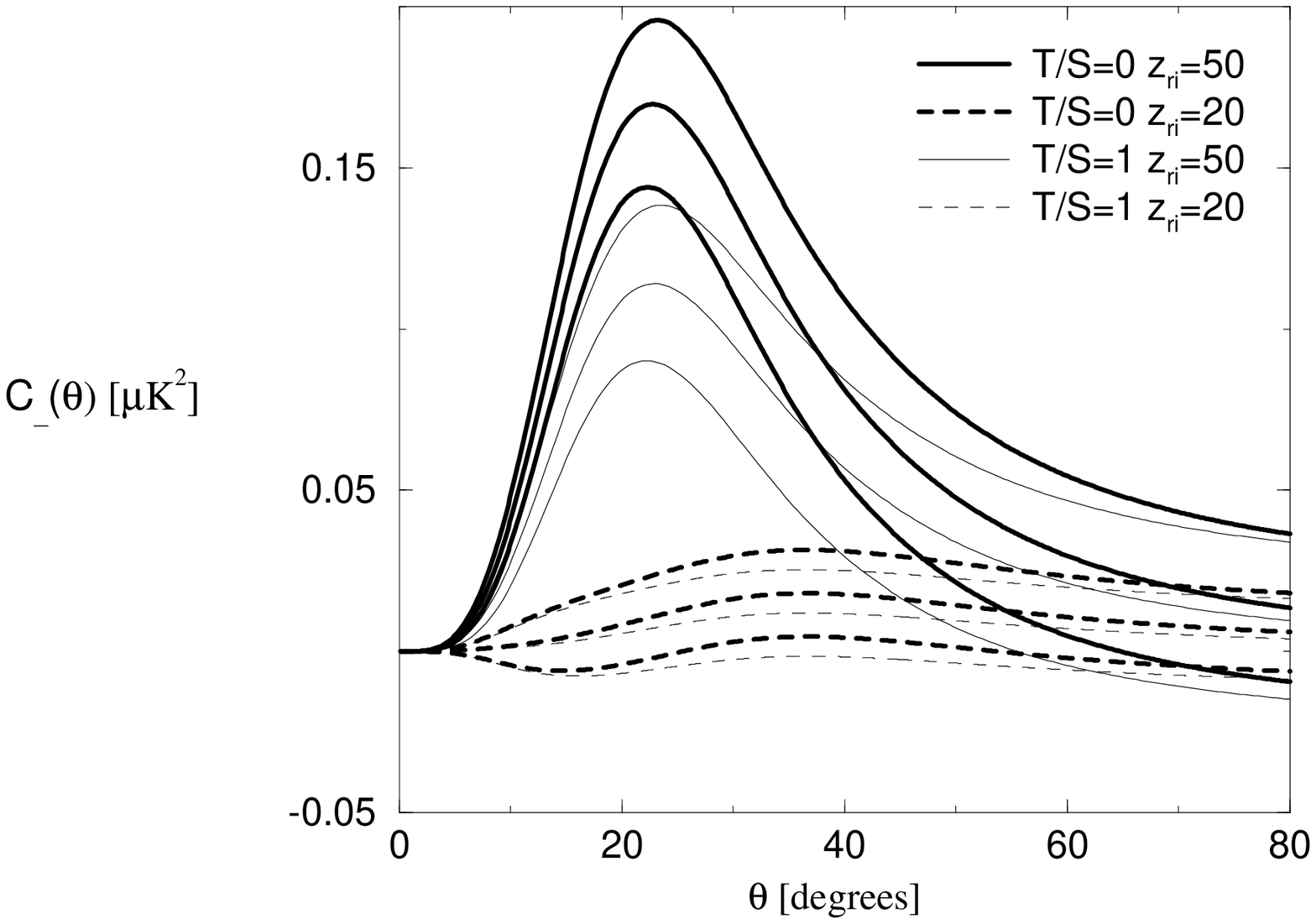}}
\caption{Full-sky averaged correlation function 
$C_-(\theta)\pm \Delta C_- (\theta)$ for SPOrt/ISS.}
\label{fig8}
\end{figure}

\begin{figure}
\leavevmode
\hbox{
\epsfxsize=6.5in
\epsffile{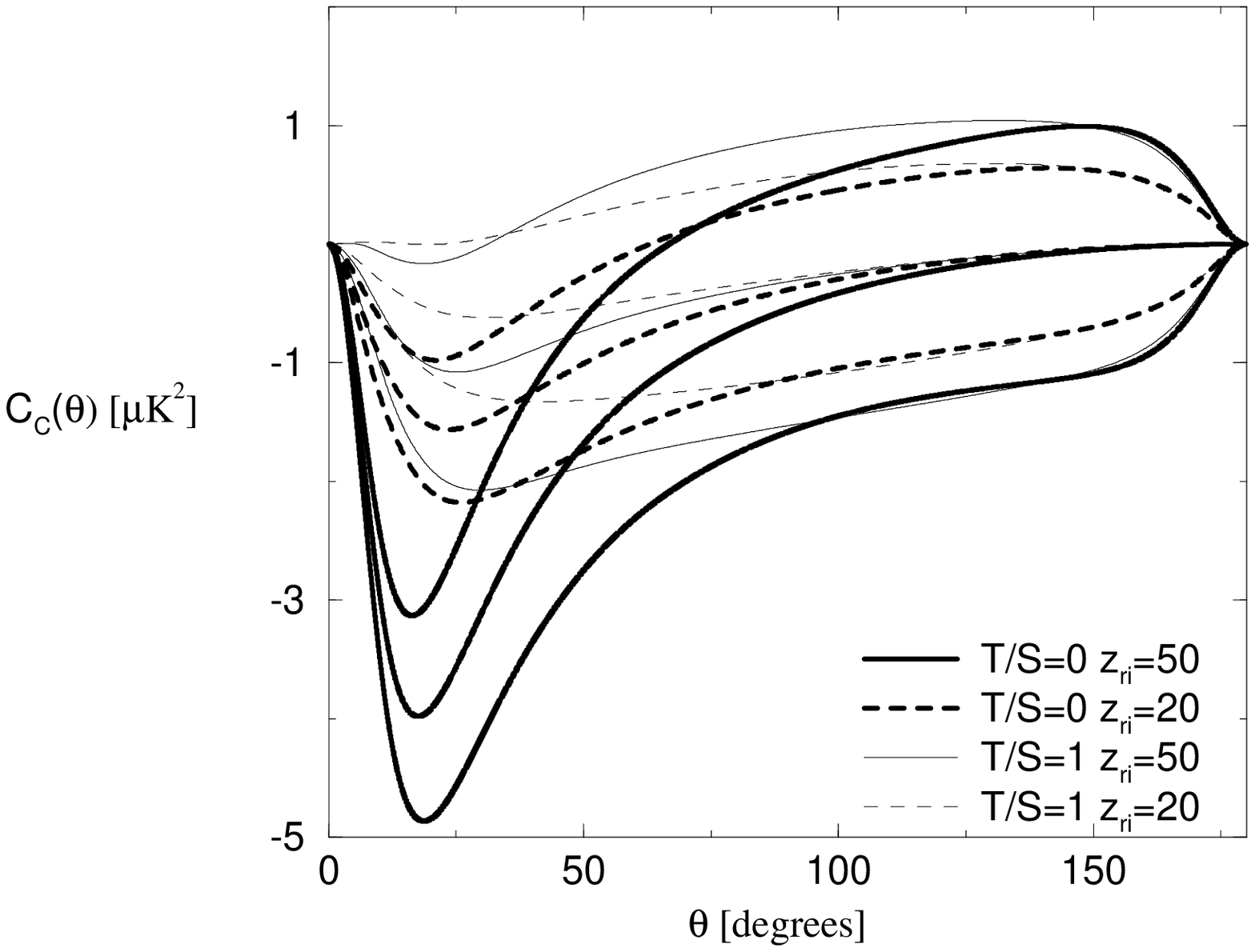}}
\caption{Full-sky averaged correlation function 
$C_C(\theta)\pm \Delta C_C (\theta)$ for SPOrt/ISS.}
\label{fig9}
\end{figure}

Another space misssion being planned is the Sky Polarization 
Observatory (SPOrt) on board the International Space Station
during the early space station utilization period (2001-2004)~\cite{cor}.
The scope is to measure the polarization of the sky diffuse background 
radiation at an angular-scale of $7^o$ for a large sky-coverage with
four frequency channels between 20 GHz and 70 GHz.
The experimental sensitivity is expected to be comparable to MAP. 
Again, we evaluated the ensemble means and variances of the full-sky averaged
correlation functions, but 
in reionized sCDM models with reionization redshifts $z_{\rm ri}=20$ and $50$. 
The results are plotted in Figs.~\ref{fig6}-\ref{fig9}.
The expected rms polarization S/N $\sim 1-3$ for $20<z_{\rm ri}<50$, 
while the anisotropy-polarization S/N $\sim 1-2$ at $\theta\sim 20^o$.

A near-term, ground-based polarization experiment, called POLAR, is to measure 
CMB polarization at $7^o$ scales for 36 pixels~\cite{kea}.
To reach a signal level of $1 \mu K$ for a single pixel
requires an integration time of about 120 hours and low-noise HEMT amplifiers
of noise temperature of about $10 K$. The expected S/N ratio of the 
rms polarization is $1-2$ for reionized sCDM models with
reionization redshifts $45<z_{\rm ri}<105$~\cite{kea}, whereas the 
anisotropy-polarization correlation would be dominated by
noise~\cite{cri1,cri2,ng4}.

\section{Likelihood Functions}\label{like}

The most straightforward way to obtain the power spectra from the measured
Stokes parameters is to perform a maximum likelihood analysis of the
data. All of the information in the measurement is encoded in the likelihood
function, which can properly take into account non-uniform detector noise, and
sample variances. This is particularly an 
advantage for ground-based experiments
which track tens or hundreds of spots in the sky to measure $Q$ and $U$.
The method offers a simple test of the consistency of the 
power spectra from map to map, and the correlation between maps.
In fact, this method has been employed by the {\it COBE}/DMR to determine
the anisotropy quadrupole normalization from the two-point functions of the
4-year anisotropy maps containing about 4000 pixels~\cite{hin}. 
However, as is known, for all-sky coverage in satellite experiments,
especially small-scale measurements, the large
amount of data involved in the computation makes the analysis inefficient. 
This problem may be reduced by using filtering and compression as in the 
case of anisotropy data. 

For a small number of measurements, such as the ongoing polarization 
experiment POLAR that measures $Q$ and $U$ by observing 
an annulus of regular $36$ spots at constant declination~\cite{kea}, 
the data set can be arranged as
\begin{equation}
{\bf D}=(Q_i+iU_i,Q_j-iU_j),
\end{equation}
where $i,j=1,..,36$. Since all data points lie on a same latitude, 
the expected theoretical correlation functions ${C_{\pm}}_{ij}$ in this case 
are given respectively by Eqs.~(\ref{c+},\ref{c-}) with $\alpha+\gamma=\pi$, i.e.
\begin{eqnarray}
{C_+}_{ij}&=&\sum_{l}\sqrt{\frac{2l+1}{4\pi}}(C_{El}+C_{Bl})
   \:_{2}W_l \:_{2}Y_{l-2}(\theta_{ij},0) e^{4i\alpha_{ij}}, \nonumber \\
{C_-}_{ij}&=&\sum_{l}\sqrt{\frac{2l+1}{4\pi}}(C_{El}-C_{Bl})
   \:_{2}W_l \:_{2}Y_{l2}(\theta_{ij},0),
\end{eqnarray}
where $\:_{2}W_l$ is the window function~(\ref{swl}) with a beam width 
appropriate to the experiment, $\theta_{ij}$ is the separation angle between 
the $i$th and $j$th spots, and $\alpha_{ij}$ (which is a geometric function
of $\theta_{ij}$) is the angle between the longitude at the $i$th spot and the 
great arc connecting the $i$th and $j$th spots (refer to Fig.~\ref{fig1}).
Thus we construct the likelihood function as
\begin{equation}
{\cal L}(C_{E2},C_{B2})=\frac{1}{\sqrt {{\rm det}{\bf C}}}
\exp\left[-{1\over 2}{\bf D}{\bf C}^{-1}{\bf D}^\dagger\right],
\end{equation}
where the correlation matrix is
\begin{equation}
{\bf C}=
\left( \begin{array}{cc} 
{C_+}_{ij} + N_{ij}& {C_-}_{ij}\\ {C_-}_{ij}& {C_+}_{ij}^* + N_{ij}
\end{array} \right),
\end{equation}
where $N_{ij}=2(\sigma^P_i)^2\delta_{ij}$ is the noise correlation matrix.
The most likely electric and magnetic quadrupoles are then determined by 
maximizing the likelihood function over the theories.

For all-sky measurements, the full likelihood function can be constructed as
\begin{equation}
{\cal L}(C_{T2},C_{C2},C_{E2},C_{B2})=\frac{1}{\sqrt{{\rm det}{\bf M}}}
\exp\left[-{1\over 2}{\bf\Delta C}{\bf M}^{-1}{\bf\Delta C}^T\right],
\end{equation}
where ${\bf M}$ is the covariance matrix of the full-sky averaged correlation
functions whose entries given by Eq.~(\ref{cm}), 
and ${\bf\Delta C}$ is a row vector with entries
\begin{equation}
\Delta C_X(\theta)\equiv C_X(\theta)_{\rm measured} - 
\left<{{\cal C}_X(\theta)}_{w_T^{-1}=w_P^{-1}=0}\right>,
\end{equation}
where the first term is the two-point correlation function in the sky map
obtained by performing the full-sky-averaging~(\ref{sky}) of  
products of all map measured Stokes parameters
with a fixed angular separation $\theta$, and the second term is calculated
from the ensemble mean of the corresponding operator listed in Eq.~(\ref{calC})
without substracting off the noise, i.e. setting $w_T^{-1}=w_P^{-1}=0$.
The tensor contribution can be analyzed by maximizing the
likelihood function with the covariance submatrix ${\bf M}_{X'Y}$, where 
$X,Y=+,-$. The central value of $C_{B2}$ in a confidence-level plot
significantly deviated from zero would indicate the presence of tensor mode.

\section{Conclusions}\label{conclusion}
It is known that the two-point correlation functions of the Stokes parameters
are explicitly dependent on coordinates. A way of getting rid of this is to
expand the Stokes parameters in terms of spin-weighted spherical harmonics,
and to construct optimal angular power spectrum estimators. 
Although being useful for all-sky coverage satellite 
experiments, it is not suitable for near-term, ground-based polarization
experiments. For a small number of observation points, the simplest way to 
compare data with theory is to perform a likelihood analysis with a correlation
function matrix. Further, a likelihood analysis of a full-sky map using 
correlation functions is a challenge for development of computational
algorithms. 

Several authors suggested to obtain coordinate-independent 
correlation functions by measuring $Q$ and $U$ with respect to axes which 
are parallel and perpendicular to the great arc connecting the two 
points being correlated~\cite{kam2}.
Here we gave the most general calculation of the two-point 
correlation functions of the Stokes parameters in terms of spin-weighted
spherical harmonics, including the windowing function and instrumental noise.
We obtained simple forms though they still explicitly depend on coordinates. 
However, the coordinate dependence can be eliminated by averaging over 
the whole sky, and the averaged correlation functions can be used to construct
the covariance matrix in likelihood analysis of future CMB satellite data.
Moreover, in ground-based polarization experiments, if a correct scanning
topology is selected, the coordinate-dependence can be eliminated or simplified
in some way and the correlation functions can be directly put in the 
correlation matrix of the likelihood function.

Furthermore, we have calculated the signal to noise ratios from the
two-point correlation functions for future anisotropy and polarization 
experiments. 
It is likely that MAP will detect the first anisotropy-polarization 
correlation signal.  In fact, to complement the MAP
measurement, a small-angle ground-based polarization experiment, targeting
at a higher signal to noise for rms polarization, should be performed as 
to cross-correlate with the MAP high-precision anisotropy map.  
Surely, measurements of the microwave sky by Planck will push cosmology 
into a new epoch, as both CMB anisotropy and polarization can be precisely
measured. On the other hand, the limit on Sunyaev-Zel'dovich distortion of
Compton-y parameter, $|y|<15\times 10^{-6}$, from FIRAS data~\cite{mat}
constrains the reionization redshift $z_{\rm ri}< 50$~\cite{bal}. This
constraint is consistent with theoretical CDM model calculations~\cite{teg},
which predict an occurrence of reionization at a redshift $30<z_{\rm ri}<70$, 
and most likely at $z_{\rm ri}\sim 50$.
So, it is probable that SPOrt/ISS would observe a polarization signal.
However, if large-scale polarization is not detected, 
then it would be an impact on cosmological theories.

\acknowledgments
This work was supported in part by the R.O.C. NSC Grant No.
NSC87-2112-M-001-039.

\newpage

\begin{tabular}{l} \hline\hline
\\$
\:_{1}Y_{2\pm2}=\pm{1\over4} \sqrt{\frac{5}{\pi}}(1\mp\cos{\theta})
      \sin{\theta}\,e^{\pm2i\phi}
$\\
$
\:_{1}Y_{3\pm2}={1\over8}\sqrt{\frac{35}{2\pi}}\left[2\sin^3{\theta}-
      \sin{\theta}(1\pm\cos{\theta})^2\right]e^{\pm2i\phi}
$\\
$
\:_{1}Y_{4\pm2}=\pm{3\over16}\sqrt{\frac{1}{2\pi}}\left[3\sin{\theta}
      (1\pm\cos{\theta})^3-5(1\mp 5\cos{\theta})\sin^3{\theta}
      \right]e^{\pm2i\phi}
$\\
$
\:_{2}Y_{2\pm2}=\frac{1}{8}\sqrt{\frac{5}{\pi}}(1\mp\cos{\theta})^2
        e^{\pm2i\phi}
$\\
$
\:_{2}Y_{3\pm2}=\pm{1\over16}\sqrt{\frac{7}{\pi}}\left[
       -(1\mp\cos{\theta})^3
       +5(1\mp\cos{\theta})\sin^2{\theta}\right]e^{\pm2i\phi}
$\\
$
\:_{2}Y_{4\pm2}={3\over32}\sqrt{\frac{1}{\pi}}\left[
       (1\mp\cos{\theta})^4-12(1\mp\cos{\theta})^2\sin^2{\theta}
       +30\sin^4{\theta}\right]e^{\pm2i\phi}
$\\
\\ \hline\hline
\end{tabular}

\bigskip
TAB. 1. Some spin-weighted spherical harmonics with $l=2,3,4$.
\newpage

\newpage

\newpage

\newpage

\newpage

\newpage

\newpage

\newpage

\newpage

\newpage

\end{document}